\documentclass[aps,prl,twocolumn,superscriptaddress,showpacs,floatfix]{revtex4}
\usepackage{graphicx}
\usepackage{dcolumn}
\usepackage{bm}
\usepackage{amsfonts}
\usepackage{amsmath}

\begin{document}
\baselineskip=0.5cm
\renewcommand{\thefigure}{\arabic{figure}}

\title{Fully Frustrated Cold Atoms}
\author{Marco Polini}
\email{m.polini@sns.it}
\affiliation{NEST-INFM and Classe di Scienze, Scuola Normale Superiore, I-56126 Pisa, Italy}
\author{Rosario Fazio}
\affiliation{NEST-INFM and Classe di Scienze, Scuola Normale Superiore, I-56126 Pisa, Italy}
\author{A.H. MacDonald}
\affiliation{Physics Department, The University of Texas at Austin, Austin, Texas 78712}
\author{M.P. Tosi}
\affiliation{NEST-INFM and Classe di Scienze, Scuola Normale Superiore, I-56126 Pisa, Italy}

\date{\today}

\begin{abstract}
Fully frustrated Josephson Junction arrays (FF-JJA's) exhibit a subtle compound phase  
transition in which an Ising transition associated with discrete broken translational 
symmetry and a Berezinskii-Kosterlitz-Thouless (BKT) transition associated with quasi-long-range phase coherence 
occur nearly simultaneously.  In this Letter we discuss a cold atom 
realization of the FF-JJA system.  We demonstrate that both orders can be studied 
by standard momentum-distribution-function measurements and present numerical results, based on 
a successful self-consistent spin-wave approximation, that illustrate the expected behavior of 
observables.   
\end{abstract} 
\maketitle

The preparation of cold atomic gases trapped in an optical lattice has opened 
up attractive new possibilities for the experimental study of strongly correlated many-particle 
systems~\cite{general_experiments} and has inspired much theoretical activity 
(see {\it e.g.} Ref.~\onlinecite{minguzzi03} for a review). In particular, the experimental 
observation by Greiner {\it et al.}~\cite{general_experiments} 
of a Superfluid-Mott Insulator (SI) transition in a three-dimensional ($3D$) optical lattice explicitly 
demonstrated the possibility of realizing strongly-correlated cold bosons.
The SI transition in an optical lattice was predicted in Ref.~\onlinecite{jaksch98} and can be described by the Bose-Hubbard model~\cite{fisher89}, which has also been employed to model $2D$ 
granular superconductors~\cite{fisher88} and JJA's~\cite{fazio01}. 
This success has motivated many new proposals~\cite{global} for cold-atom simulations 
of strongly correlated boson phenomena.

In this Letter we propose that cold atoms be used to study the incompletely understood 
phase transitions that occur in FF-JJA's~\cite{fazio01,frustration}. 
The boson Hubbard model for JJA's accounts for Cooper 
pair hopping between small superconducting particles and for Coulomb interactions which 
can be dominantly intra-particle.  For superconducting particles the model applies when the thermal energy $k_BT$ 
is much smaller than the bulk energy gap, {\it i.e.} when the underlying fermionic character of 
electrons is suppressed.  Cold atoms in optical lattice potentials provide, in some senses
at least, a closer realization~\cite{jaksch98,general_experiments} of the boson Hubbard model
because other degrees of freedom are more completely suppressed and because the interactions 
are more dominantly on-site. Frustration~\cite{frustration} 
can be introduced into JJA's by introducing an external magnetic field to change 
the energetically preferred phase relationship between boson amplitudes on neighboring sites.
Frustration in this case refers to the impossibility of choosing the optimal phase difference for 
each bond.  In a cold atom optical lattice system, frustration can be introduced 
by altering the phase factors for atom hopping between optical potential minima more 
explicitly, for example by following procedures similar to those proposed recently 
by Jaksch and Zoller~\cite{zollerjjafield}, Mueller~\cite{mueller}, and S\o rensen {\it et al}.~\cite{demlerlukinjjafield}.
The laser configurations suggested in these papers also enable spatially periodic modulation of the magnitude of 
boson hopping amplitudes, a feature that is important to the proposal outlined below.

\begin{figure}
\includegraphics[scale=0.40]{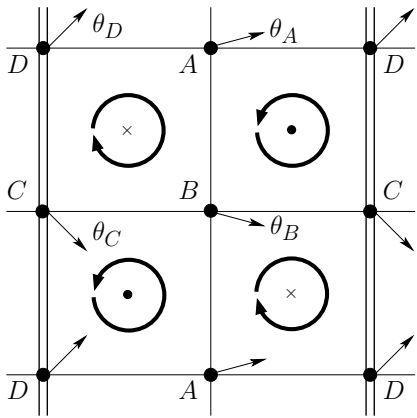}
\caption{Ground-state degenerate solutions for a classical FF-JJA. Double vertical lines stand for modulated ``antiferromagnetic" bonds ($-\alpha\,E_J$) while single vertical and horizontal lines stand for unmodulated ``ferromagnetic bonds" ($E_J$). The configuration shown corresponds to $\alpha=0.5$ for which $\theta_A=-\theta_B=\pi/12$ and $\theta_D=-\theta_C=\pi/4$. The distinct configuration with equal energy is obtained by $(\theta_A,\theta_D)\rightarrow-(\theta_A,\theta_D)$ or equivalently by vertical translation by one lattice constant.\label{fig:one}}
\end{figure}

In a FF square-lattice JJA
the sum of the optimal 
phase differences for individual bonds around every plaquette is $\pi$, {\em fully} 
incompatible with the integer multiple of $2\pi$ phase winding constraint imposed by the 
single-valued condensate wavefunction.
For square lattice JJA's full frustration can be introduced by applying an 
external magnetic field that generates one half of a superconducting flux quantum through
each plaquette of the array.  In the Landau gauge the frustration is imposed by changing 
the sign of every second vertical hopping parameter. 
For a FF-JJA, the Gross-Pitaevskii mean-field 
equation of the corresponding boson Hubbard model has  
two distinct degenerate solutions, illustrated schematically in Fig.~\ref{fig:one},
which break the discrete translational symmetry of the lattice, and for each solution a free overall phase factor in the condensate wavefunction which breaks gauge symmetry. 
The surprising property of FF square lattice JJA's, and by extension of 
FF square lattice cold atoms, is that the Ising order and the quasi-long-range 
phase order appear to vanish nearly simultaneously and continuously at a common critical temperature.
When quantum fluctuations are included, similar phase changes are expected to occur at zero temperature
as the on-site interaction strength is increased.  If these orders do in fact 
disappear simultaneously, the phase transition would have to be in a new universality class and 
could not have a natural description in terms of the condensate wavefunction order-parameter,
a situation reminiscent of the deconfined quantum critical behavior discussed
recently~\cite{senthil04} by Senthil {\it et al}.

The compound phase change in a frustrated JJA is closely related to the phase 
changes that occur in the vortex lattices of the mixed state of superconductors,
and in rotating $^4{\rm He}$ and cold atom systems~\cite{cornell,rotatingcoldatomtheory}. 
The vortex lattice ground state has broken translational symmetry instigated by 
frustrating order-parameter-phase dependent terms in the Hamiltonian.
The key difference between vortex lattices and frustrated JJA's is that
the broken translational symmetry is discrete rather than continuous in the latter case.
Thermal fluctuations of 
a vortex lattice imply~\cite{moore1} that quasi-long-range phase 
order cannot exist at any finite temperature in $2D$ systems. 
For superconductors it has been argued~\cite{moore2} that given the 
absence of phase coherence, broken translational symmetry will 
not occur either.  For the FF-JJA case,
the opposite conclusion has been reached in a careful Monte Carlo study by Olsson~\cite{olsson};
he finds that that vortex position fluctuations suppress the phase stiffness and instigate a 
BKT transition as the Ising phase transition temperature is approached 
from below.  If correct, this conclusion would have to be altered when frustration
is weakened, as described below, and the Ising transition temperature is driven to zero.
In this Letter we point out that these subtle phase changes can be studied 
by measuring the momentum distribution function (MDF) of a FF cold atom cloud,
and report on theoretical estimates for the MDF based on a self-consistent harmonic
approximation (SCHA)~\cite{ariosa}.

We assume that atom hopping between sites on the optical 
lattice is weak enough to justify a single-band Wannier basis~\cite{jaksch98} 
with Wannier function $w({\bf x})$. The lattice Hamiltonian we study is
\begin{equation}\label{eq:mqpm}
{\hat{\cal H}}_f=\frac{U}{2}\sum_{{\bf x}_i}
{\hat n}^2_{{\bf x}_i}-\sum_{{\bf x}_i, {\bm \delta}} E^J_{{\bf x}_i,{\bm \delta}}
\cos{({\hat \phi}_{{\bf x}_i}-{\hat \phi}_{{\bf x}_i+{\bm \delta}})}
\end{equation}
where ${\bf x}_i=d(n,m)$ with $n,m\in [-{\cal N},{\cal N})$ is on a 
$2D$ square lattice with lattice constant $d$, ${\bm \delta}$ is the vector connecting a lattice site to its neighbours, 
and the Josephson energy or atom hopping energies
$E^J_{{\bf x}_i,{\bm \delta}}$ are identical (equal to $E_J$) on all bonds except the vertical 
bonds on every
second column.  These modulated frustrating bonds have the value $-\alpha E_J$ with $\alpha > 0$~\cite{ariosa}. 
In Eq.~(\ref{eq:mqpm}) the phase operator ${\hat \phi}_{{\bf x}_i}$ has been introduced by approximating
the atom annihilation operator on site ${\bf x}_i$ by    
$
{\hat b}_{{\bf x}_i}\simeq \sqrt{{\bar n}}\exp{(i {\hat \phi}_{{\bf x}_i})}
$, 
allowed when the mean occupation ${\bar n}$ on each lattice site is large. 
The density ${\hat n}_{{\bf x}_i}$ and phase ${\hat \phi}_{{\bf x}_i}$ operators are canonically conjugate on each site. The negative hopping parameters introduce frustration, which can be energetically weakened~\cite{ariosa} 
by choosing $\alpha < 1$. 
 
When quantum fluctuations are neglected, the $T=0$ condensate phase pattern~\cite{frustration} 
is determined by minimizing the classical energy with respect to the phase difference $\chi$ across positive $E_J$ links; the single-valued condition requires that the magnitude of the phase difference across negative $E_J$ links 
$\chi'=-3\chi$, implying~\cite{frustration} that $\sin(\chi)=\alpha \sin(3\chi)$ and hence that 
\begin{equation}\label{eq:chi}
\chi=\pm\arcsin{(\sqrt{[(3\alpha-1)/\alpha]}/2)}
\end{equation}
for $\alpha > 1/3$, while $\chi=0$ for $\alpha < 1/3$.  For $\alpha <  1/3$, the energy penalty of frustration 
is paid completely on the negative $E_J$ link and the classical ground ground state condensate 
phase is spatially constant.  As $\alpha$ increases beyond this value, the energy penalty of frustration is
increasingly shifted to the positive $E_J$ links.  The ground state configuration in this regime is doubly
degenerate with currents circulating in opposite directions around alternating plaquettes, as illustrated in 
Fig.~\ref{fig:one}. Thermal and quantum fluctuations will degrade both Ising and phase coherence orders.  
   
Phase coherence of cold atoms in an optical lattice can be directly 
detected by observing a multiple matter-wave interference pattern after ballistic expansion 
with all trapping potentials switched off. As time evolves, phase-coherent matter waves
that are emitted from each lattice site overlap and interfere with each other. 
Narrow peaks appear in the MDF due 
a combination of lattice periodicity and long-range phase coherence~\cite{pitaevskii01,cuccoli01,roth03}.
The vortex superlattice of the $\alpha > 1/3$ mean-field state 
results in the appearance of additional peaks in the MDF;
$n_f({\bf k})=\Re e\langle {\hat \Psi}^{\dagger}({\bf k}) {\hat \Psi}({\bf k})\rangle/A$ 
where $A$ is the system area, and ${\hat \Psi}({\bf k})$ is the $2D$ Fourier transform
of the field operator,
$
{\hat \Psi}({\bf x})=\sum_{{\bf x}_i} w({\bf x}-{\bf x}_i) {\hat b}_{{\bf x}_i}
$. 
It follows that 
\begin{equation}\label{eq:momentumdistribution}
n_f({\bf k})= \frac{{\bar n} |w({\bf k})|^2}{A} 
\Re e\sum_{{\bf x}_i,{\bf x}_j}e^{i {\bf k}\cdot ({\bf x}_i-{\bf x}_j)}\, C({\bf x}_i,{\bf x}_j)
\end{equation}
where we have defined a Wannier function form factor 
$w({\bf k})=\int d^2{\bf x}\,e^{-i{\bf k}\cdot{\bf x}} w({\bf x})$ and 
the phase-phase correlator $C({\bf x}_i,{\bf x}_j) \equiv \langle 
\exp{[i({\hat \phi}_{{\bf x}_i}-{\hat \phi}_{{\bf x}_j})]}\rangle$.
In the broken translation symmetry state 
$n_f({\bf k})$ is non-zero at superlattice reciprocal lattice vectors ${\bf G}_{n,m} = \pi (n,m)/d$; 
for the classical ({\it i.e.} $U=0$) ground state at zero temperature we find that 
$n_f({\bf G})=(N^2_{\rm s}/A) {\bar n} |w({\bf G})|^2 S_0({\bf G})$ where $N_{\rm s}=4{\cal N}^2$ is the total number of lattice sites, and the superlattice structure factors are   
\begin{eqnarray} \label{eq:sk_0} 
S_0({\bf G}_{0,0}) &=& [\cos{(\chi)}\cos{(\chi/2)}]^2\nonumber \\ 
S_0({\bf G}_{1,0}) &=& [\sin{(\chi)}\sin{(\chi/2)}]^2\nonumber \\ 
S_0({\bf G}_{0,1}) &=& [\sin{(\chi)}\cos{(\chi/2)}]^2\nonumber \\
S_0({\bf G}_{1,1}) &=& [\cos{(\chi)}\sin{(\chi/2)}]^2 
\end{eqnarray} 
with $S_0({\bf G}_{n+2k,m+2k})=S_0({\bf G}_{n,m})$ for any integers $n$, $m$, and $k$.  
Phase coherence in a lattice leads to condensation peaks in
$n_f({\bf k})$ at all reciprocal lattice vectors ${\bf G}_{2n,2m}$.  Coherence 
and Ising broken translational symmetry leads to additional peaks (satellites) 
with the characteristic pattern of structure factors summarized 
by Eqs.~(\ref{eq:sk_0}) at the $2 \times 2$ superlattice reciprocal lattice vectors.
MDF measurements therefore probe both types of order.

These results will be altered by both quantum and thermal fluctuations. At low 
temperature ($k_B T \ll E_J$) and well inside the superfluid 
regime ($U \ll E_J$), the phase correlation functions are 
given reliably by a SCHA~\cite{ariosa} in which the density matrix is approximated 
by that of an effective harmonic model defined by  
mean condensate phases on each site and harmonic 
coupling constants $K$ on each nearest neighbour link.  
Minimizing the variational free-energy with respect to mean phases
enforces average current conservation at each node of the lattice. Minimization with 
respect to the harmonic coupling constants sets them equal to the self-consistently
determined mean curvature of the Josephson interaction.  The phase changes
across the vertical and horizontal positive $E_J$ links, $\theta_{h}$ and $\theta_{v}$, are 
unequal in this approximation, as are the harmonic coupling constants $K_{h}$ and $K_{v}$
and (of course) the coupling constant on frustrated links $K_{\alpha}$.  
For $U \rightarrow 0$ and $T \rightarrow 0$, the $\theta_{h}=\theta_{v} \rightarrow \chi$,
$K_{h}=K_{v} \rightarrow E_{J}\cos{\chi}$ and $K_{\alpha} \to -\alpha E_{J}\cos{(3\chi)}$.

The SCHA phase correlation function 
$C({\bf x}_i,{\bf x}_j)=C^{\mu,\nu}_{\rm NF}C^{\mu,\nu}_{\rm Q}({\bf X}_i,{\bf X}_j)$ 
is the product of a long-range factor 
$C^{\mu,\nu}_{\rm NF}$, dependent only on position within the $2 \times 2$ broken-symmetry 
unit cell, and a Gaussian factor $C^{\mu,\nu}_{\rm Q}({\bf X}_i,{\bf X}_j)$
which captures the power-law decay of phase correlations in $2D$ superfluids 
(here ${\bf X}_i$ is a lattice vector of the large unit cell so that
sites are labelled by $\mu$ and $i$). We find that $C^{\mu,\nu}_{\rm NF}$ is given by 
\begin{equation}\label{eq:cnf}
C^{\mu,\nu}_{\rm NF}=
\left(
\begin{array}{cccc}
1&e^{i\theta_v}&e^{i(\theta_v+\theta_h)}&e^{-i\theta_h}\\
e^{-i\theta_v}&1&e^{i\theta_h}&e^{-i(\theta_h+\theta_v)}\\
e^{-i(\theta_v+\theta_h)}&e^{-i\theta_h}&1&e^{-i(\theta_v+2\theta_h)}\\
e^{i\theta_h}&e^{i(\theta_h+\theta_v)}&e^{i(\theta_v+2\theta_h)}&1
\end{array}
\right)\,,
\end{equation}
and that
\begin{eqnarray}\label{eq:c_quantal}
C^{\mu,\nu}_{\rm Q}({\bf X}_i,{\bf X}_j)&=&
\exp
\biggl\{
-\frac{U}{N^2_{\rm s}}
\sum_{\sigma}
\sum_{{\bf k} \in {\rm BZ}'}
\frac{{\cal F}^{\mu,\nu}_{{\bf k},\sigma}({\bf X}_i-{\bf X}_j)}{\xi_{{\bf k},\sigma}}
\biggr.
\nonumber \\
&\times&
\biggl.
[1+2\,N_{\rm BE}(\xi_{{\bf k},\sigma}/k_BT)]
\biggr
\}\,.
\end{eqnarray}
In Eq.~(\ref{eq:c_quantal}) the sum is over the four Bogoliubov eigenmodes of the harmonic Josephson term 
at each wavevector in the $2 \times 2$ super cell's Brillouin zone.
Because we have chosen strictly on-site interactions, the quantum harmonic problem
can be solved by first diagonalizing the Josephson interaction term,
as in the classical case, and then performing independent Bogoliubov
transformations on each mode. The contribution of a given Bogoliubov mode to the mean 
square phase difference between sites $(\mu,i)$ and $(\nu,j)$ in Eq.~(\ref{eq:c_quantal}) 
is therefore characterized by the quantity~\cite{ariosa}
\begin{eqnarray}\label{eq:formfactor}
{\cal F}^{\mu,\nu}_{{\bf k},\sigma}({\bf X}_i-{\bf X}_j)&=&
|v^{\sigma}_{\mu}({\bf k})|^2+|v^{\sigma}_{\nu}({\bf k})|^2-2\Re e \Bigl\{[v^{\sigma}_{\mu}({\bf k})]^\star \Bigr.\nonumber\\
&\times&\Bigl.v^{\sigma}_{\nu}({\bf k})e^{i{\bf k}\cdot ({\bf X}_i-{\bf X}_j+{\bf b}_{\mu\nu})}\,\Bigr\}\, 
\end{eqnarray}
where ${\bf b}_{\mu\nu}$ is the site separation for $i=j$, $N_{\rm BE}(x)$ is a Bose-Einstein thermal factor, $\xi^2_{{\bf k},\sigma}=U\lambda_{{\bf k},\sigma}$, $\lambda_{{\bf k},\sigma}$ and $v^{\sigma}_{\mu}({\bf k})$ being the eigenvalues and the $\mu$-th component of the eigenvectors of the harmonic Josephson interaction.  

\begin{figure}
\includegraphics[scale=0.60]{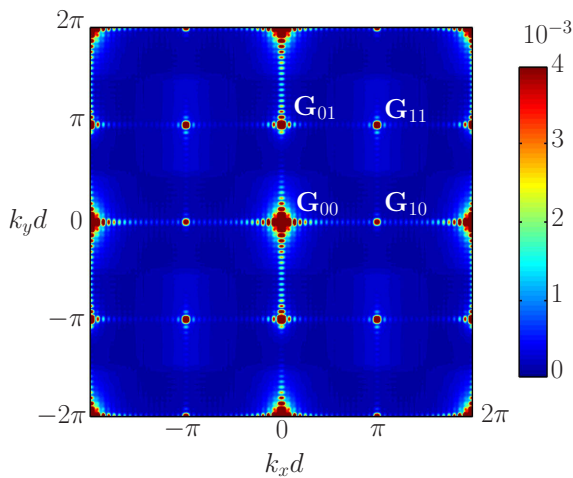}
\caption{The structure factor $S({\bf k})$ for FF cold bosons in a $2D$ array with $\alpha=0.5$ 
as a function of the continuous variable ${\bf k}d \in [-2\pi,2\pi]\times[-2\pi,2\pi]$. Here $T=0.242E_J/k_B$~\cite{footnote}, and $U=0.1 E_J$.\label{fig:two}}
\end{figure}

We have evaluated $S({\bf k})=n_f({\bf k})A/({\bar n}N^2_{\rm s}|w({\bf k})|^2)$ in the presence of both quantum and thermal fluctuations by summing over a finite lattice with $N_{\rm s}=1296$ sites in Eq.~(\ref{eq:momentumdistribution}) and applying periodic boundary conditions to make the wavevectors in Eq.~(\ref{eq:c_quantal}) discrete.  A typical result is reported in Fig.~\ref{fig:two}. The presence of non-zero Ising satellites at ${\bf k}={\bf G}_{1,0}, {\bf G}_{0,1}$ and ${\bf G}_{1,1}$ is evident. 
These peaks are a sharp manifestation of the broken discrete translational symmetry and would be absent in an unfrustrated system.

\begin{figure}
\includegraphics[scale=0.40]{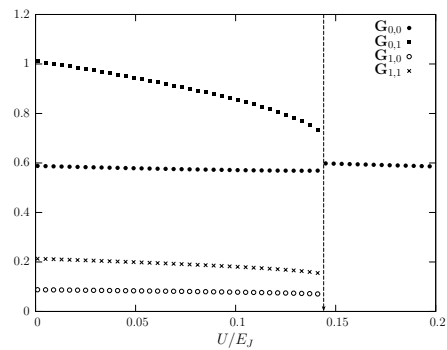}
\caption{Condensation and Ising peaks of the structure factor $S({\bf G})$ as a function of $U/E_J$. The value of $S({\bf G})$ for ${\bf G}_{1,0}$, ${\bf G}_{0,1}$ and ${\bf G}_{1,1}$ has been multiplied by a factor of $10$ for clarity. The vertical dashed line indicates the value of $U^{c}_{\rm IS}$.\label{fig:three}}
\end{figure}

The evolution of $S({\bf k})$ with $U$ at fixed $T=0.242\,E_J/k_B$ is illustrated in Fig.~\ref{fig:three}
where we plot $S({\bf G})$ for ${\bf G}_{0,0}$, ${\bf G}_{1,0}$, ${\bf G}_{0,1}$ and ${\bf G}_{1,1}$. 
All four peaks are slightly suppressed by quantum and thermal fluctuations with respect to the $U=T=0$ values in Eq.~(\ref{eq:sk_0}). At the critical value $U^{c}_{\rm IS} \approx 0.14$ the Ising satellites disappear while the condensation peak survives (the first 
order character of this transition is an artifact of the SCHA). 
The superlattice peaks may be regarded as Ising order parameters $\sim {\cal S}=\sin{(\theta_h)}$ (see Eq.~(\ref{eq:sk_0})). At the Ising point $\theta_h \rightarrow 0$, causing $S({\bf G}_{0,0})$ to increase 
with increasing $U$, before resuming its decline. 

In summary, we have shown that FF cold atoms can offer a unique opportunity for experimental study of 
a system in which there is competition between critical phenomena associated with ${\mathbb Z}_2$ and gauge ${\rm U}(1)$ broken symmetries.

\acknowledgments 
This work was partially supported by EC-RTN ``Nanoscale Dynamics" and
by an Advanced Research Initiative of S.N.S. 
A.H.M. acknowledges support from the 
Welch Foundation and from the National Science Foundation 
under grant DMR-0115947. We gratefully acknowledge 
the early contributions to this work by Jairo Sinova. 
We wish to thank R. Asgari, P. Capuzzi, B. Davoudi, M. Gattobigio, M. Greiner, and K. Madison for useful discussions.

\newpage

\begin{thebibliography}{99}
\bibitem{footnote}
	This value of $T$ has been chosen to be slightly below the Ising critical temperature $T^c_{\rm IS}$ for $U=0$ 
	just for computational convenience. In fact the only requirement on $T$ 
	to observe the Ising peaks in a MDF measurament is $T<T^c_{\rm IS}$. As $T$ decreases, the critical value 
	$U^c_{\rm IS}$ for an Ising transition in the presence of interactions rapidly increases.
\end{thebibliography}
\end{document}